# High-performance and Low-power Transistors Based on Anisotropic Monolayer β-TeO$_2$

Shiying Guo,[1,2] Hengze Qu,[1] Wenhan Zhou,[1] Shengyuan A Yang,[2,*] Yee Sin Ang,[3] Jing Lu,[4] Haibo Zeng,[1,†] Shengli Zhang[1,‡]

[1]MIIT Key Laboratory of Advanced Display Materials and Devices, School of Materials Science and Engineering, Nanjing University of Science and Technology, Nanjing 210094, China

[2]Research Laboratory for Quantum Materials, Singapore University of Technology and Design, Singapore 487372, Singapore

[3]Science, Mathematics and Technology, Singapore University of Technology and Design, Singapore, Singapore

[4]State Key Laboratory of Mesoscopic Physics, Department of Physics, Peking University, Beijing, 100871, People's Republic of China

E-mail:
[*]shengyuan_yang@sutd.edu.sg
[†]zeng.haibo@njust.edu.cn
[‡]zhangslvip@njust.edu.cn

**Abstract**:

Two-dimensional (2D) semiconductors offer a promising prospect for high-performance and energy-efficient devices especially in the sub-10 nm regime. Inspired by the successful fabrication of 2D β-TeO$_2$ and the high on/off ratio and high air-stability of fabricated field effect transistors (FETs) [Nat. Electron. 2021, 4, 277], we provide a comprehensive investigation of the electronic structure of monolayer β-TeO$_2$ and the device performance of sub-10 nm metal oxide semiconductors FETs (MOSFETs) based on this material. The anisotropic electronic structure of monolayer β-TeO$_2$ plays a critical role in the anisotropy of transport properties for MOSFETs. We show that the 5.2-nm gate-length n-type MOSFET holds an ultra-high on-state current exceeding 3700 μA/μm according to International Roadmap for Devices and Systems (IRDS) 2020 goals for high-performance devices, which is benefited by the highly anisotropic electron effective mass. Moreover, monolayer β-TeO$_2$ MOSFETs can fulfill the IRDS 2020 goals for both high-performance and low-power devices in terms of on-state current, sub-threshold swing, delay time, and power-delay product. This study unveils monolayer β-TeO$_2$ as a promising candidate for ultra-scaled devices in future nanoelectronics.

# I. INTRODUCTION

The continuous down-scaling demand of metal-oxide-semiconductor field-effect transistors (MOSFETs) inspires the intensive exploration of novel channel semiconductors in the past decades [1]. Two-dimensional (2D) semiconductors exhibit excellent electrostatic gate control ability and is capable of suppressing carrier scattering even though in the sub-10 nm regime because of their atomic-scale thickness and dangling bond-free smooth surface [2-5]. Therefore, 2D semiconductors are expected to hold strong advantages as the candidate channel materials for ultra-scaled MOSFETs. Motivated by the promising potential of 2D semiconductors, tremendous research efforts have been devoted to exploring the incorporation of 2D semiconductors as the channel of MOSFETs. Recent experiments have successfully demonstrated MOSFET operations of multiple 2D semiconductors, including $MoS_2$, InSe, $Bi_2O_2Se$, tellurene [6-18]. In addition, computational device modeling of sub-10-nm MOSFETs has also been carried out intensively in recent decades in order to search for 2D semiconductors and device architecture with exceptional performance figures of merit [19-24].

Among the emerging 2D semiconductors, 2D black phosphorene MOSFETs have been demonstrated to exhibit an on/off ratio on the order of $10^5$ and the charge-carrier mobility can reach up to 1000 $cm^2V^{-1}s^{-1}$ [25-29]. Previous theoretical researches suggested that when the channel length is shortened to sub-10 nm, the high on-state currents (~4000 $cm^2V^{-1}s^{-1}$) and fast switching speed can also be achieved in black phosphorene MOSFETs [30,31]. Benefiting from the highly anisotropic band structure, black phosphorene MOSFETs are capable of delivering better device performance when compared to other 2D materials with nearly isotropic electronic properties, such as $MoS_2$ [32-34]. The anisotropic effective mass directly leads to the generation of high saturation current with steep sub-threshold swing (SS), thus realizing a high on-state current [35,36]. However, black phosphorene is unstable in ambient atmosphere and the performance of MOSFETs suffers from severe degeneration. Therefore, the search of anisotropic 2D semiconductors with high stability and high device performance is an urgent task for the development of next generation nanoelectronics.

2D beta tellurium dioxide (β-TeO$_2$) is predicted to be an oxide semiconductor with highly anisotropic crystal and electronic structures with a wide band-gap [37]. Recently, 2D β-TeO$_2$ nanosheets have been successfully synthesized through the surface oxidation of a eutectic mixture of tellurium and selenium [38]. The synthesized bilayer β-TeO$_2$ has a band-gap of 3.7 eV, indicating its optical transparency within the visible light spectrum. In addition, the fabricated β-TeO$_2$ MOSFET is highly stable under ambient condition and exhibits a low SS and an on/off ratio exceeding $10^6$. The hole carrier mobility reaches 232 cm$^2$ V$^{-1}$ s$^{-1}$ at room temperature and 6000 cm$^2$ V$^{-1}$ s$^{-1}$ at −50 °C. Despite the physical properties of 2D β-TeO$_2$ have been closely scrutinized, the transport properties of 2D β-TeO$_2$, especially its compatibility for ultra-scaled nanoelectronics and the device performance limit of 2D β-TeO$_2$ MOSFETs, remain an open question thus far.

In this work, we investigate the electronic structure and transport properties of monolayer β-TeO$_2$ by combining the density functional theory (DFT) with nonequilibrium Green's function (NEGF) method. The electronic transport properties of sub-10 nm gate-length MOSFETs are found to be highly anisotropic due to the anisotropic electronic structure of monolayer β-TeO$_2$. We demonstrate that n-type monolayer β-TeO$_2$ MOSFET with 5.2 nm gate length can sustain an ultra-high on-state current up to 3750 μA/μm due to the low electron effective mass along y direction, thus fulfilling the International Roadmap for Devices and Systems (IRDS) 2020 goals for high-performance (HP) devices [39]. The proposed monolayer β-TeO$_2$ MOSFETs can be further scaled down to 4.0 nm for HP applications and 5.2 nm for low-power (LP) applications, thus suggesting the versatility of 2D β-TeO$_2$ in meeting the demands of next-generation nanoelectronics. In addition, the on-state current, delay time and power consumption are predicted to exceed most of the other 2D-material-based MOSFETs. These findings reveal the potential of monolayer β-TeO$_2$ MOSFETs as a strong contender in nanoelectronics, with outstanding performance in both the HP and LP device specifications of future technology nodes.

## II. COMPUTATIONAL METHODS

The geometry optimization and electronic properties are carried out using the DFT package in the Atomistix ToolKit (ATK) 2020 version [40]. The DFT calculations are performed by the generalized gradient approximation in the form of Perdew-Burke-Ernzerhof realization using PseudoDojo potentials [41]. Monkhorst-Pack k-point mesh is used, with size 25 × 25 × 1 for geometric optimization and 33 × 33 × 1 for electronic property calculation [42]. The density mesh cut-off is set to 75 Hartree.

The device performance is simulated by combining the DFT and NEGF method in ATK 2020. The drain current $I_{ds}$ is calculated based on the Landauer-Bűttiker formula [43,44]. $I_{ds}$ at a given bias voltage $V_b$ and gate voltage $V_g$ is obtained by

$$I_{ds}(V_b, V_g) = \frac{2e}{h} \int_{-\infty}^{+\infty} \{T(E, V_b, V_g)[f_s(E - \mu_s) - f_d(E - \mu_d)]\}dE \quad (1)$$

where $T(E, V_b, V_g)$, $f_{s/d}$, and $\mu_{s/d}$ are the transmission function, the Fermi-Dirac distribution functions for the source/drain, and the electrochemical potentials of the source/drain, respectively. And the difference value between $\mu_s$ and $\mu_d$ is equal to $e \times V_b$. The k-dependent transmission coefficient $T_{k_{//}}(E)$ is obtained by

$$T_{k_{//}}(E) = \text{Tr}\{\Gamma^s_{k_{//}}(E) G_{k_{//}}(E) \Gamma^d_{k_{//}}(E)[G_{k_{//}}(E)]^\dagger\} \quad (2)$$

where $\Gamma^{s/d}_{k_{//}}(E) = i[\Sigma^{s/d}_{k_{//}} - (\Sigma^{s/d}_{k_{//}})^\dagger]$ represents the broadening width deriving from source/drain in the form of self-energy $\Sigma^{s/d}_{k_{//}}$. $G_{k_{//}}(E)$ and $[G_{k_{//}}(E)]^\dagger$ are the retard and advanced Green's function, respectively. The reciprocal lattice vector $k_{//}$ is perpendicular to the transport direction. The average of k-dependent transmission coefficient leads to the transmission function $T(E)$.

## III. RESULTS AND DISCUSSION

We first discuss the optimized crystal structure and electronic properties of monolayer β-TeO$_2$, including band structure and effective mass. The anisotropic atomic structure of monolayer β-TeO$_2$ is shown in Fig. 1(a), with the optimized lattice parameters of $a$ = 5.44 Å and $b$ = 5.76 Å. Monolayer β-TeO$_2$ possesses P2$_1$/c symmetry

(No. 14), containing four tellurium atoms and eight oxygen atoms in each unit cell. One tellurium atom is bonded to four neighboring oxygen atoms while one oxygen atom is bonded to two neighboring tellurium atoms. Figure 1(a) also shows the anisotropic atomic structure of monolayer β-TeO$_2$, which reveals the possible anisotropy of its electronic properties.

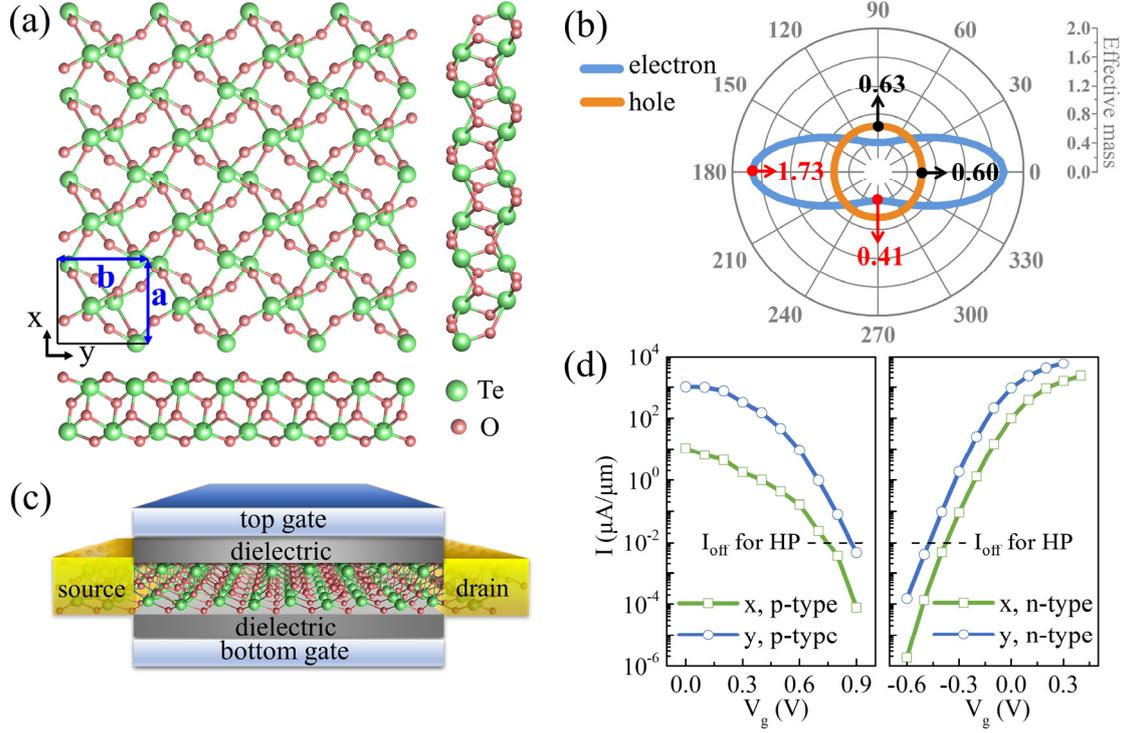

FIG. 1. (a) Top and two side views of monolayer β-TeO$_2$ structure. (b) Effective mass for electrons and holes of monolayer β-TeO$_2$. 0° and 90° angles represent the x and y directions, respectively. (c) Schematic illustration of the double-gated monolayer β-TeO$_2$ MOSFET. The dielectric constant of dielectric layer is ε = 4.0. The channel length is equal to the gate length ($L_g$). (d) Transfer characteristics of the p- and n-type monolayer β-TeO$_2$ MOSFETs, with $L_g$ = 5.4 nm for the x direction and $L_g$ = 5.2 nm for the y direction. The supply voltage is $V_{ds}$ = 0.65 V.

The band structure and effective mass of monolayer β-TeO$_2$ are then calculated, as shown in Fig. S1 within the Supplemental Material and Fig. 1(b), respectively. The band-gap is found to be 2.59 eV for monolayer β-TeO$_2$. The valence band maximum

(VBM) is located at Γ point, while the conduction band minimum (CBM) is 0.06 × 2π/x away from Γ point, which is very close to the Γ point. The monolayer β-TeO$_2$ is thus a nearly direct band-gap semiconductor. For the VBM, the band dispersions of monolayer β-TeO$_2$ along the x and y directions are almost identical. However, the band dispersions around the CBM along y direction are difficult to be determined directly from the band structure as shown in Fig. S1 within the Supplemental Material. In order to clear illustrate the energy dispersions, we calculate the effective mass for electrons and holes along different directions in x-y plane, as shown in Fig. 1(b). It can be clearly seen that the hole effective mass is nearly isotropic with 0.60 m$_0$ and 0.63 m$_0$ along the x direction and y direction, respectively. In contrast, the electron effective mass along the x direction is 1.73 m$_0$, which is significantly larger than the 0.41 m$_0$ along the y direction.

Table I. The hole (h) and electron (e) effective mass of monolayer β-TeO$_2$, and the on-state current ($I_{on}$) of 5-nm gate-length monolayer β-TeO$_2$ MOSFETs along different transport directions.

|  | x, h | y, h | x, e | y, e |
| --- | --- | --- | --- | --- |
| Effective mass (m$_0$) | 0.60 | 0.63 | 1.73 | 0.41 |
|  | x, p-type | y, p-type | x, n-type | y, n-type |
| $I_{on}$ (μA/μm) | 7 | 615 | 1300 | 3750 |

We next explore the device performance of p- and n-type monolayer β-TeO$_2$ MOSFETs, along x and y transport directions. The schematic drawing of the simulated MOSFET is shown in Fig. 1(c). The monolayer β-TeO$_2$ channel is sandwiched between top and bottom dielectric layers and metal gates. Highly doped monolayer β-TeO$_2$ is adopted as the source and drain of the device and it is assumed that the external metal electrodes form efficient Ohmic contacts with monolayer β-TeO$_2$ without the need of thermal charge injection over a contact Schottky barrier [45]. Different doping concentrations are investigated to optimize the device performance (see Figs. S2 and

S3 within the Supplemental Material). The optimized electrode doping concentration for the n-type MOSFET along the y direction is $6 \times 10^{13}$ cm$^{-2}$, while the optimized doping concentration for others is $8 \times 10^{13}$ cm$^{-2}$.

On-state current ($I_{on}$) is an important figure of merit that can be acquired from the transfer characteristics, which is evaluated at $V_{g, on} = V_{ds} + V_{g, off}$, where $V_{g, on}$ and $V_{g, off}$ represent the gate voltage of the on-state and off-state, respectively, and $V_{ds}$ is the supply voltage which is equal to the $V_b$. The transfer characteristics of p- and n-type monolayer β-TeO$_2$ MOSFETs with 5 nm gate length along x and y directions are shown in Fig. 1(d). The off-state current ($I_{off}$) is taken to be 0.01 μA/μm according to the IRDS 2020 goals for HP devices in 2028, marked as black dashed line in Fig. 1(d). The obtained $I_{on}$ is summarized in Table I.

For the n-type MOSFETs with 5 nm gate length, the on-state current along the y direction reaches 3750 μA/μm, which is much higher than that along the x direction (1300 μA/μm), since the anisotropic electron effective mass is 0.41 $m_0$ along the y direction versus 1.73 $m_0$ along the x direction. The difference between the on-state currents along different transport directions can also be identified from the transmission spectrum in the on-state (see Fig. S4 within the Supplemental Material). On the other hand, the unexpected on-state currents are observed for the p-type MOSFETs along the x and y directions. Although the hole effective mass in the VBM exhibits an isotropic characteristic, the on-state current along the x direction (7 μA/μm) is two orders of magnitude smaller than that along the y direction (615 μA/μm).

In order to understand the connection between the electronic band structure and device performance, the local band structure around the CBM and VBM of monolayer β-TeO$_2$ and the corresponding transmission coefficients of MOSFETs are computed, as shown in Figs. 2 and 3. As shown in Figs. 2(a) and 2(b), the energy dispersions of CBM along the x direction show a slight "M" shape with the minimum point located at X" point, while the conduction band along y direction shows a sharper peak at X" point. The transmission coefficients along two transport directions in Figs. 2(c) and 2(d) reveal that the shape and the width of the on-state transmission coefficient window are sensitively dependent on the band dispersion near the CBM. When the transport direction is along the x direction, the transmission coefficient versus transverse wave vector $k_y$ in Fig. 2(c) has one hill, which is similar to the band dispersions along the y direction. When the transport is along the y direction, the transmission coefficient versus the transverse wave vector $k_x$ exhibits an "M" shape profile, which is akin to the band dispersions along the x direction. Furthermore, the transmission coefficient versus $k_x$ spans over a wider wave vector window than that versus $k_y$, thus resulting in a higher on-state transport current when the MOSFET is operated along the y direction.

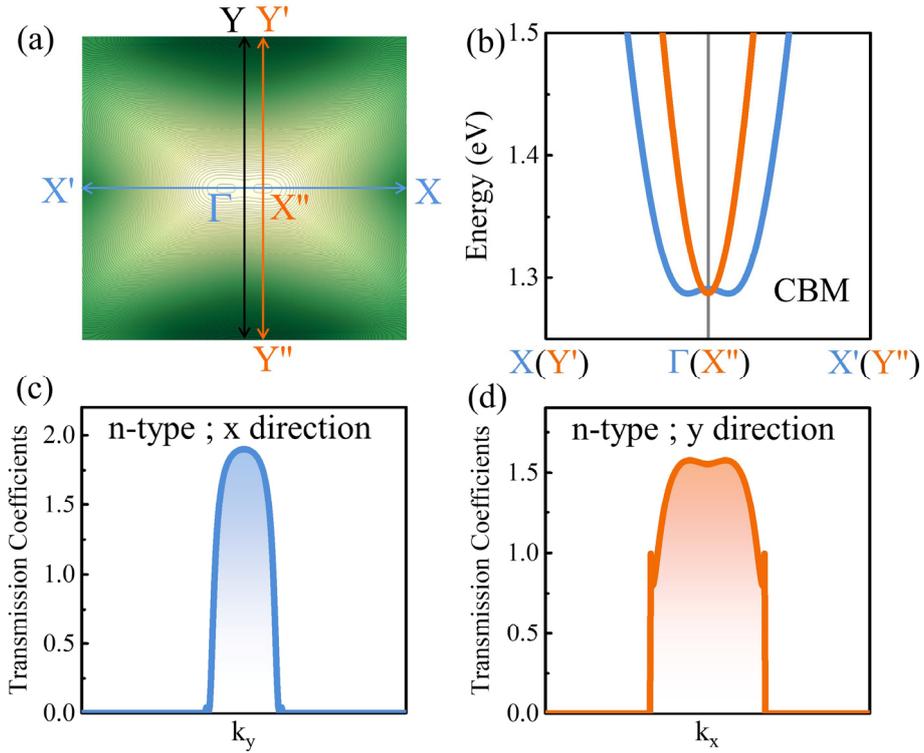

FIG. 2. The (a) energy contour plots of the lowest conduction band and (b) local conduction band around the Γ point in first Brillouin zone of monolayer β-TeO$_2$. The transmission coefficients versus transverse wave vector (c) $k_y$ and (d) $k_x$ of 5-nm gate-length n-type monolayer β-TeO$_2$ MOSFETs with the transport direction along the x and y directions in the on-state, respectively.

For electron transport mediated by the valence band of monolayer β-TeO$_2$, there is only one valley near VBM as located at the Γ point, as shown in Figs. 3(a) and (b). The band dispersions along the x and y directions are almost the same at the energy level larger than −1.45 eV. However, when the energy level is lower than −1.45 eV, a suddenly weaker band dispersion appears along the x direction, which is in stark contrast to that along the y direction. Correspondingly, the transmission coefficient versus $k_y$ with x direction as the transport direction is significantly suppressed in comparison with that versus $k_x$ with y direction as the transport direction in Fig. 3(c). Such contrasting transmission coefficient windows arise from the suddenly appeared flat band dispersion at −1.45 eV and explains the extremely low on-state current (only 7 μA/μm) of the x-directed MOSFET.

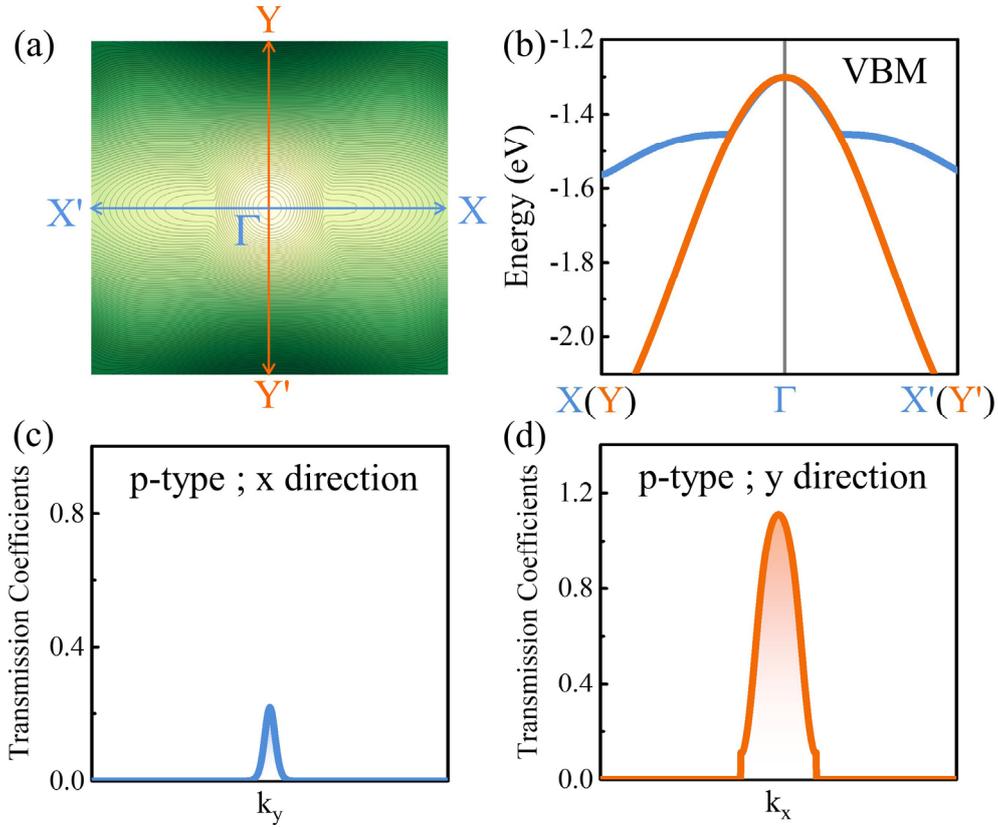

FIG. 3. The (a) energy contour plots of the highest valence band and (b) local valence band around the Γ point in first Brillouin zone of monolayer β-TeO$_2$. The transmission coefficients versus transverse wave vector (c) $k_y$ and (d) $k_x$ of 5-nm gate-length p-type monolayer β-TeO$_2$ MOSFETs with the transport direction along the x and y directions in the on-state, respectively.

To further verify whether the holes at the −1.45 eV energy level contribute to the device performance, we investigate the valence band structure, density of states (DOS), and hole concentration of high-doped monolayer β-TeO$_2$, as shown in Fig. S5 within the Supplemental Material. The doping concentration is set to 8 × 10$^{13}$ cm$^{-2}$ because the injected carriers are provided by the highly-doped monolayer β-TeO$_2$ source region. After the p-type doping, the VBM shifts up to 0.09 eV and the Fermi level lies below the valence band. At thermal equilibrium, the energy distribution function for holes is obtained by the Fermi-Dirac distribution $f_h(E)$ [46],

$$f_h(E) = 1 - \frac{1}{\exp\left(\frac{E - E_F}{kT}\right) + 1} \tag{3}$$

where $E_F$, $k$ and $T$ represent the Fermi energy level, Boltzmann's constant, room temperature, respectively. The hole concentration is defined as

$$p(E) = D_h(E) f_h(E) \tag{4}$$

Where $D_h(E)$ represents the DOS for holes. The calculated hole concentration for highly-doped monolayer β-TeO$_2$ is shown in Fig. S5(c) within the Supplemental Material. As the holes around −0.07 eV at the flat band near VBM is within the energy level range of the hole concentration, these holes with a large effective mass thus have a substantial impact on the device performance and has resulted in extremely low transmission coefficients and on-state current in the x-directed p-type MOSFET, as compared to the y-directed p-type MOSFET.

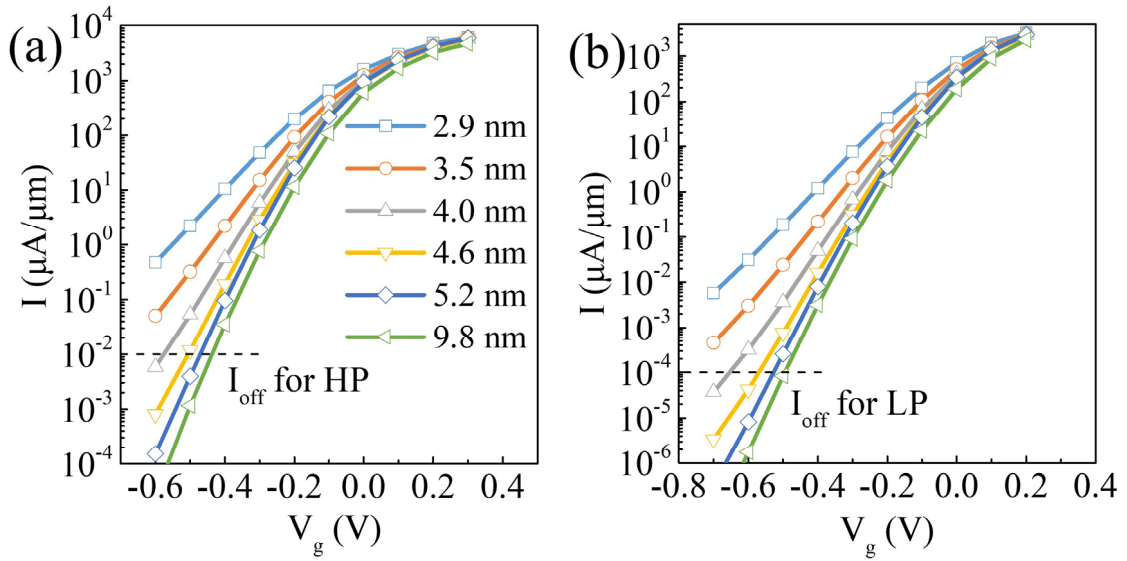

FIG. 4. Transfer characteristics of n-type monolayer β-TeO$_2$ MOSFETs along the y

direction for (a) HP and (b) LP requirements according to IRDS 2020. The off-state current for HP and LP requirements is marked as black dashed line in (a) and (b), respectively.

Considering that only the n-type monolayer β-TeO$_2$ MOSFETs along the y direction can fulfill the IRDS 2020 requirements, we further investigate the n-type configuration along the y direction with other sub-10-nm channel lengths to explore its performance limit. We first focus on HP devices. The electrode doping concentration is the same as that of 5.2 nm n-type monolayer β-TeO$_2$ MOSFET. The transfer characteristics are presented in Fig. 4(a) and the calculated figures of merit are summarized in Table SI within the Supplemental Material. The leakage current increases with the decreasing channel length. When the channel is shorter than 4.0 nm, the leakage current is overly high and fails to meet the HP requirements due to the severe short channel effect. Hence, the channel length of 4.0 nm can be regarded as the theoretical limit for monolayer β-TeO$_2$ MOSFET with an on-state current of 2010 μA/μm. Notably, the MOSFETs with the channel length larger than 5.2 nm possess an on-state current higher than 3400 μA/μm.

Subthreshold swing (SS) is another key figure of merit to evaluate the gate electrostatic control ability of logic devices, which is defined as the gate voltage required to change the current by one order of magnitude. For the channel lengths ranging from 9.8 nm to 5.2 nm, the SS (i.e. 67~72 mV/dec) is close to the thermionic limit of 60 mV/dec and fulfills the IRDS 2020 requirements for HP devices in the 2028 horizon (i.e. SS < 75 mV/dec). The SS increases slowly with a shorter channel. When the channel is shorter than 5.2 nm, SS rises rapidly with a shorter channel, reaching 86 mV/dec for 4.6 nm channel and 98 mV/dec for 4.0 nm channel.

Besides, we also compare the device performances of 2D β-TeO$_2$ with other 2D materials based on the standards of the International Technology Roadmap for Semiconductors (ITRS) 2013 [47]. According to the ITRS 2013 standards where the off-state current is fixed at 0.1 μA/μm, we summarize the important figures of merit such as the on-state current and SS in Table SII within Supplemental Material. Due to the higher off-state current standard of ITRS 2013, the on-state current of the n-type monolayer β-TeO$_2$ MOSFETs increases by more than 1000 μA/μm for the same channel length and the channel limit is shortened to 3.5 nm. The highest on-state current reaches 5140 μA/μm in the 5.2 nm gate-length monolayer β-TeO$_2$ MOSFETs. As shown in Fig. 5(a), for sub-10 nm nodes, monolayer β-TeO$_2$ MOSFETs possess a significantly higher on-state current than the MOSFETs based on other 2D materials, such as black phosphorene and Bi$_2$O$_2$Se [48-51], thus revealing the strong switchability of monolayer β-TeO$_2$ for logic devices.

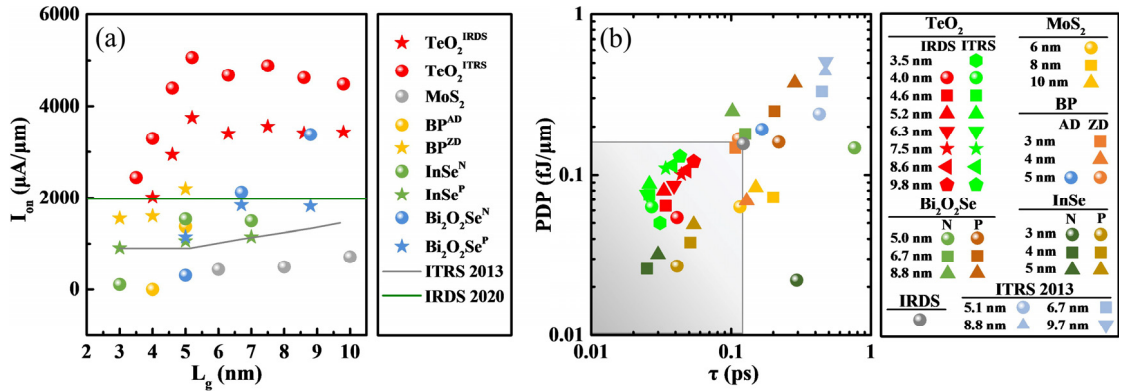

FIG. 5. The (a) on-state current ($I_{on}$), and (b) power-delay product (PDP) versus delay time (τ) of n-type monolayer β-TeO$_2$ MOSFETs along y direction for HP requirements according to two different standards, IRDS 2020 for the year 2028 and ITRS 2013. The data for MOSFETs based on other 2D materials according ITRS 2013 are shown for comparison [48-51]. The superscript "AD" and "ZD" represent that the transport directions of MOSFETs are along armchair and zigzag directions, respectively. The superscript "N" and "P" represent the n-type and p-type MOSFETs, respectively.

Apart from the on-state current, the intrinsic gate delay time (τ) and the power-delay product (PDP) are another two key indicators for transistors. The intrinsic gate delay time τ describes the upper limit of the switching speed,

$$\tau = \frac{(Q_{on} - Q_{off})}{I_{on}} \quad (5)$$

where $Q_{on}$ and $Q_{off}$ are the charges in the channel region in the on- and off-state, respectively. PDP is the power consumption in one switching operation and is

calculated by

$$\text{PDP} = V_{ds}I_{on}\tau = V_{ds}(Q_{on} - Q_{off}) \tag{6}$$

The PDP versus $\tau$ of n-type monolayer β-TeO$_2$ MOSFETs along y direction for HP requirements is shown in Fig. 5(b). According to IRDS 2020, sub-10 nm monolayer β-TeO$_2$ MOSFETs have a series of delay times ranging from 0.041 ps to 0.054 ps and PDP ranging from 0.054 fJ/μm to 0.121 fJ/μm. These data are all satisfy the HP standards of IRDS 2020 in 2028 horizon, as denoted in the blue shaded region in Fig. 5(b). We then compare the $\tau$ and PDP with other 2D material MOSFETs. The standards are adopted in accordance with the ITRS 2013. The $\tau$ and PDP of monolayer β-TeO$_2$ MOSFETs are lower than those of the MOSFETs based on MoS$_2$, black phosphorene, Bi$_2$O$_2$Se. In addition, monolayer β-TeO$_2$ MOSFETs have a little higher PDP than 2D InSe MOSFETs, while their delay time are comparable. Overall, these results indicate the compelling potential of the monolayer β-TeO$_2$ MOSFETs as a channel material for ultra-scaled HP MOSFETs.

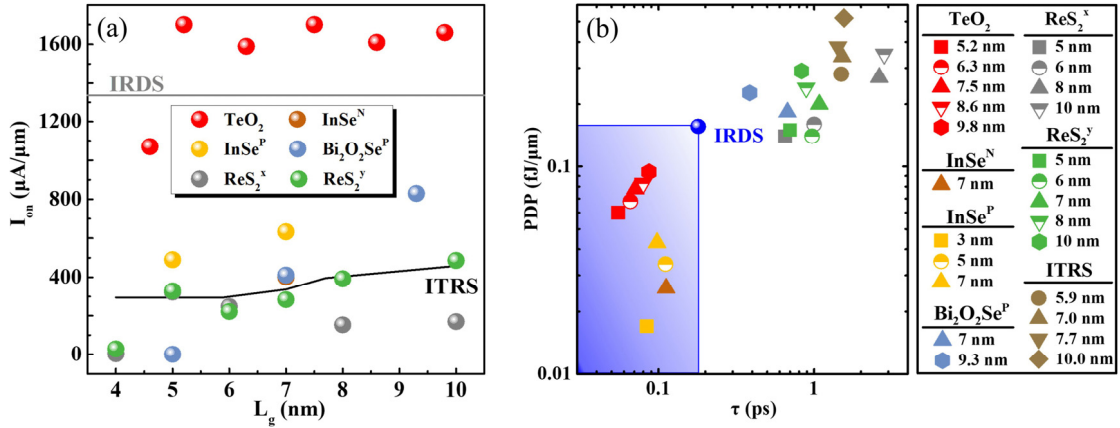

FIG. 6. The (a) on-state current ($I_{on}$), and (b) power-delay product (PDP) versus delay time ($\tau$) of n-type monolayer β-TeO$_2$ MOSFETs along y direction for LP requirements according to IRDS 2020 for the year 2028. The data for MOSFETs based on other 2D materials according ITRS 2013 are shown for comparison [49,50,52]. The superscript "x" and "y" represent that the transport directions of MOSFETs are along x and y directions, respectively.

In contrast to HP devices, LP devices require a lower off-state current to serve the purpose of saving static energy. For LP device, the optimized doping concentration of electrodes for LP transistors is $4 \times 10^{13}$ cm$^{-2}$. Figure 4(b) shows the transfer characteristics of LP monolayer β-TeO$_2$ MOSFETs under different channel lengths. The on-state current for LP devices is evaluated with off-state current fixed at $1 \times 10^{-4}$ μA/μm according to the IRDS requirements. The MOSFETs in the cases of L$_g$ > 5 nm provides the higher on-state current than that in the IRDS requirements (see Table SIII within the Supplemental Material). While HP transistors are more dependent on the superthreshold characteristics, the LP transistors rely more on the subthreshold characteristics [53]. LP monolayer β-TeO$_2$ MOSFETs display lower SS than HP, ranging from 67 to 64 mV/dec for 5.2~9.8 nm channel. The small values of SS indicate the good device electrostatics of LP monolayer β-TeO$_2$ MOSFETs. Furthermore, the on-state currents of LP monolayer β-TeO$_2$ MOSFETs are higher than those of other 2D semiconductor LP MOSFETs as shown in Fig. 6(a), indicating a higher on-off ratio and operating speed.

We next evaluate whether the $\tau$ and PDP of n-type monolayer β-TeO$_2$ MOSFETs can satisfy the LP requirements of IRDS 2020. The calculated $\tau$ and PDP are shown in Fig. 6(b) and Table SIII within the Supplemental Material. The $\tau$ and PDP gradually decrease as the channel length shortens, which are consistent with previous studies [52]. The $\tau$ (0.055~0.087 ps) and PDP (0.060~0.094 fJ/μm) fulfill the LP requirements of IRDS 2020 for all channel lengths down to 5.2 nm. In particular, the $\tau$ and PDP of monolayer β-TeO$_2$ MOSFETs are far lower than those of ReS$_2$ and Bi$_2$O$_2$Se MOSFETs in the sub-10 nm nodes. Compared with InSe MOSFETs, monolayer β-TeO$_2$ MOSFETs have a higher PDP but a shorter $\tau$. The short intrinsic delay time and low power consumption thus further reveals the promising potential of monolayer β-TeO$_2$ MOSFETs for LP applications in digital circuits.

## IV. CONCLUSIONS

In summary, we explore the electronic properties of monolayer β-TeO$_2$ and performance limits of sub-10 nm gate-length monolayer β-TeO$_2$ MOSFETs. Our results demonstrate that monolayer β-TeO$_2$ is a promising candidate for HP and LP devices. The anisotropic band dispersions near VBM and CBM lead to the anisotropic device performance of monolayer β-TeO$_2$ MOSFETs. The performance of the n-type monolayer β-TeO$_2$ MOSFETs along y direction meets the IRDS 2020 requirement in terms of on-state current, SS, delay time and PDP. In particular, the on-state current can reach up to 3750 μA/μm due to the corresponding anisotropic effective mass. Combining its advantage of high air stability, monolayer β-TeO$_2$ is therefore a competitive air-stable channel material in the application of nanoelectronics and transparent logic devices.

## ACKNOWLEDGEMENTS

This work was financially supported by the Training Program of the Major Research Plan of the National Natural Science Foundation of China (91964103), the Natural Science Foundation of Jiangsu Province (BK20180071), the Fundamental Research Funds for the Central Universities (No.30919011109), and also sponsored by Qing Lan Project of Jiangsu Province, the Six Talent Peaks Project of Jiangsu Province (Grant No. XCL-035), and the Singapore MOE AcRF Tier 2 (MOE-T2EP50220-0011).